\documentclass{aastex631}

\usepackage{amsmath}
\usepackage{graphicx}
\usepackage{natbib}

\NewPageAfterKeywords

\shorttitle{Companion to Betelgeuse}

\shortauthors{Howell, Ciardi, Clark, et al.}

\begin{document}

\title{The Probable Direct-Imaging Detection of the Stellar Companion to Betelgeuse}

\correspondingauthor{Steve B. Howell}
\email{steve.b.howell@nasa.gov}

\author[0000-0002-2532-2853]{Steve~B.~Howell}
\affiliation{NASA Ames Research Center, 
Moffett Field, CA 94035 USA}

\author[0000-0002-5741-3047]{David R.~Ciardi}
\affil{NASA Exoplanet Science Institute-Caltech/IPAC, Pasadena, CA 91125, USA}

\author[0000-0002-9144-7726]{Catherine A. Clark}
\affil{NASA Exoplanet Science Institute-Caltech/IPAC, Pasadena, CA 91125, USA}

\author[0000-0002-9144-7726]{Douglas A. Hope}
\affiliation{Georgia Tech Research Institute, 925 Dalney Street, Atlanta, GA, 30318, USA
and Georgia State University, 25 Park Place, Atlanta, GA 30303, USA}

\author[0000-0002-9144-7726]{Colin Littlefield}
\affiliation{NASA Ames Research Center, 
Moffett Field, CA 94035 USA and Bay Area Environmental Research Institute, Moffett Field, CA 94035, USA}

\author[0000-0001-9800-6248]{Elise Furlan}
\affil{NASA Exoplanet Science Institute-Caltech/IPAC, Pasadena, CA 91125, USA}

\begin{abstract}
Betelgeuse -- the closest M-supergiant to the Sun -- has recently been predicted to host a lower-mass stellar companion that orbits the primary with a period of $\sim 6$ years. The putative stellar companion is thought to cause the long photometric modulation observed in Betelgeuse, which cannot be explained by stellar pulsations.
Additionally, radial velocity and astrometric data also point to a stellar companion. Here we present diffraction-limited optical speckle imaging observations obtained on the 8.1-meter Gemini North telescope in 2020 and 2024. The 2020 observations were taken during the Great Dimming event, and at a time when the stellar companion was predicted to be unobservable because it was directly in-line with Betelgeuse itself. The 2024 observations were taken three days after the predicted time of greatest elongation for the companion. A comparison of the 2020 and 2024 data reveal no companion in 2020 (as expected) and the probable detection of a companion in 2024. The presumed stellar companion has an angular separation and position angle of 52 mas and $115^\circ$ east of north, respectively, which is in excellent agreement with predictions from dynamical considerations. The detected companion is roughly six magnitudes fainter than Betelgeuse at 466 nm. While this is only a 1.5$\sigma$ detection, the results are in reasonable agreement with the predictions: the appearance of the companion at quadrature; the angular separation from Betelgeuse; the position angle with respect to Betelgeuse; the magnitude difference; and the estimated mass of the companion.
\end{abstract}


\section{Introduction \label{sec:Intro}}

Betelgeuse has been the target of observational studies for centuries. Its location (visible from both hemispheres), its brightness \citep[the tenth-brightest visually and the brightest in the infrared;][]{Johnson1966CoLPL...4...99J}, and its prominent red color have made it a landmark in the sky and throughout history.

Betelgeuse was the first star, other than the Sun, to have a direct measurement of its angular diameter \citep[47 mas at 575 nm;][]{mp1921}, and observations at a variety of wavelengths have yielded similar results: 37 mas at 535 nm \citep{RoddierRoddier1983ApJ...270L..23R}, 54 mas in a red optical TiO absorption band \citep{wilson1992MNRAS.257..369W}, and 43 mas in the $K$-band \citep{Perrin2004A&A...418..675P, Montarges2016A&A...588A.130M}. Betelgeuse was also the first star, other than the Sun, to be directly imaged using the Hubble Space Telescope
\citep{1996ApJ...463L..29G}.
Variations in the angular size of Betelgeuse have been observed and attributed to surface brightness distributions, as Betelgeuse pulsates over an approximately 400-day cycle \citep{RoddierRoddier1983ApJ...270L..23R, Joyce2020ApJ...902...63J}. The effective angular size varies with wavelength due to line absorption/emission \citep{wilson1992MNRAS.257..369W,Perrin2004A&A...418..675P}. The latest stellar models of Betelgeuse adopt a photospheric angular diameter of $\sim42$ mas, with multiple shells and asymmetries extended to 50-60 mas \citep{Kervella2018A&A...609A..67K}, and ultraviolet emission arising from chromospheric ions extending to $\sim135$ mas \citep{UU1998AJ....116.2501U}. Due to this large angular diameter, Betelgeuse can be resolved using a number of telescopes and instrument combinations, if it does not saturate the detector.

Variations in the visual brightness of Betelgeuse have also been observed; this variability was first described in \citet{1836nbvc.book.....H}, and many detailed photometric and spectroscopic studies of Betelgeuse have been performed in the nearly two hundred years since \citep[see the extended discussion in][]{MAC2025ApJ...978...50M}. These studies have demonstrated that the star shows complex variability in brightness and radial velocity due to convection and pulsation.

\begin{figure*}[h]
\centering
\includegraphics[width=\textwidth]{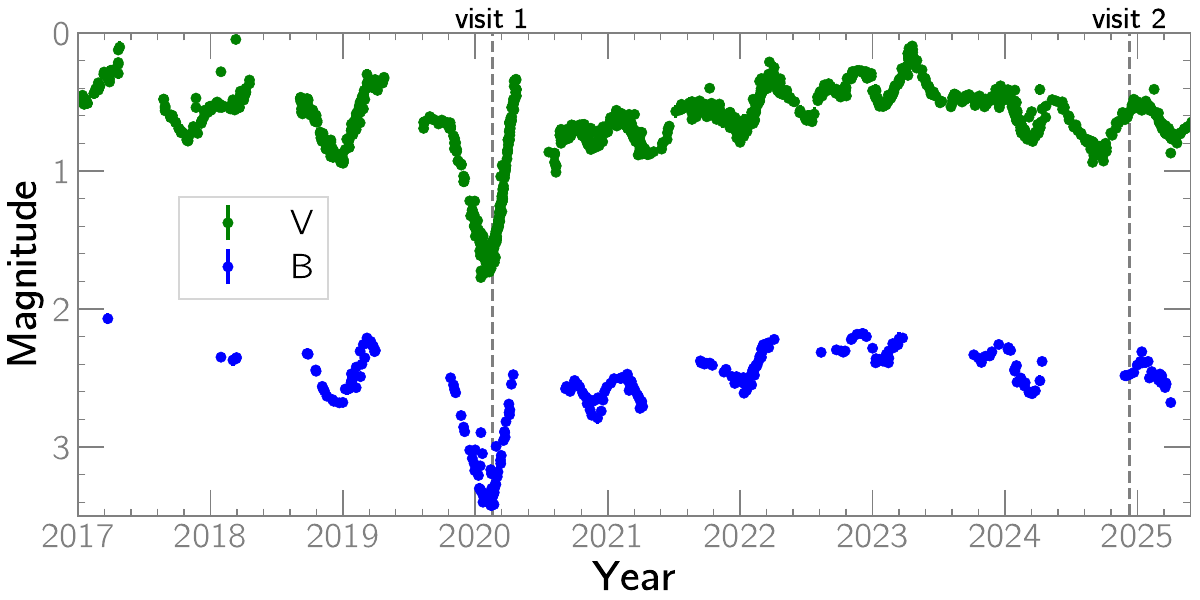}
\caption{The American Association of Variable Star Observers (AAVSO) photometric observations of Betelgeuse. The $V$ and $B$ band measurements cover the past eight years including the deep fading of Betelgeuse in late 2019 to early 2020 known as the Great Dimming event. The dates of our two speckle observations are marked. 
}
\label{fig:B-LC}
\end{figure*}

A large photometric drop in brightness, known as the Great Dimming event, transpired in 2019-2020, when the brightness of Betelgeuse faded by $V\sim1.6$ mag (Figure \ref{fig:B-LC}). While some initially thought this was evidence for an upcoming supernova, this event was later determined to be evidence of a large surface mass ejection that led to dust production in the atmosphere \citep{Montarges2021Natur.594..365M, dupree2022ApJ...936...18D, MacLeod2023ApJ...956...27M}.

Another persistent variation is a $\sim6$-year secondary period visible in photometric, radial velocity, and astrometric measurements of Betelgeuse. As such, many studies have proposed the presence of a companion (or companions) to Betelgeuse, and the reader is referred to the discussions and references listed in several recent and superbly detailed papers on the star \citep{Gold2024ApJ...977...35G, MAC2025ApJ...978...50M, Goldberg2025arXiv250518375G, OGrady2025arXiv250518376O}. Of interest here is that these recent exhaustive works, using a wealth of data covering over a century, not only come to the conclusion that Betelgeuse does indeed have a close companion, but also derive similar properties for the companion star. The suspected companion star has an orbital period near 6 years and a nearly edge-on orbit to our line-of-sight. They propose that the companion is the cause of the long secondary period (LSP) photometric variations and the astrometric and radial velocity variations. However, direct observational detection of the companion via normal imaging was deemed unlikely because of the large predicted brightness difference between Betelgeuse and its companion, as well as the overall brightness of Betelgeuse itself.

Spurred on by the Great Dimming event in 2020, and later by the recent works advocating for the existence of a stellar companion to Betelgeuse, we undertook high-resolution optical speckle imaging observations using the 8.1-meter Gemini North telescope. The large aperture of the Gemini telescope, the optical wavelengths at which we observed, and the fast readout capabilities of the speckle camera detectors provided the best opportunity to directly observe Betelgeuse with sufficient angular resolution to measure the angular diameter of Betelgeuse and search for its proposed companion, without saturating Betelgeuse itself.

We present here the results from our speckle observations taken in 2020 (when the companion was predicted to be unobservable) and those taken in 2024 (when the companion was predicted to be at its greatest elongation), and report the likely direct-imaging detection of a stellar companion to Betelgeuse.

In Section \ref{sec:methods}, we report on our observations and data reduction methodologies. In Section \ref{sec:results}, we present angular diameter measurements for Betelgeuse and the probable direct-imaging detection of its stellar companion. In Section \ref{sec:discussion}, we discuss the likelihood that our detection is real and bound, as well as the nature of the companion. In Section \ref{sec:summary}, we summarize our findings and consider future work.

\section{Observations}
 \label{sec:methods}

We obtained speckle images of Betelgeuse in February 2020 and December 2024. These observations were made using the 'Alopeke speckle imager \citep{scott:2021} on the 8.1-meter Gemini North telescope in Hawai'i. ‘Alopeke provides high-resolution speckle imaging simultaneously in two optical bandpasses using two Andor Electron-Multiplying Charge-Coupled Device cameras. Within 'Alopeke, the collimated beam of light from the telescope is split by a dichroic at 700 nm to produce a blue and a red channel, each with a variety of narrow- and broad-band filters. For each target, thousands of very short (10-60 millisecond) exposures are collected to ``freeze" the atmospheric distortion in each exposure \citep{fried:1966}, as this is the rate at which an isoplanatic patch of atmosphere varies with time. The many thousands of short exposures are then mathematically combined using Fourier analysis techniques as described in \citet{howell:2011, horch:2011a} to produce diffraction-limited reconstructed optical images. During each observing run, several calibration binaries with varying separations, position angles, and magnitude differences are observed. These binaries have well-known orbital elements and are used to assess the plate scales and orientations of the two cameras. The plate scales were 0.009522 arcsec/pixel 
in February 2020 and 
0.009352 arcsec/pixel in December 2024.

Our first observation of Betelgeuse occurred on UT 2020 February 17, during the Great Dimming event. We observed Betelgeuse at 562 nm and obtained 6,000 60-millisecond exposures. The point spread function (PSF) standard star HR 2075 was used to calibrate all observations. Betelgeuse had a $V$ magnitude of 1.4 mag at the time of these observations (Figure \ref{fig:B-LC}).

Our second observation of Betelgeuse occurred on UT 2024 December 09. During these observations, the exposure time was only 10 milliseconds, as Betelgeuse was brighter ($V\sim0.6$ mag). We observed Betelgeuse at 466 nm (with a 40-nm bandpass), which is near the central wavelength of the standard $B$ filter (445 nm). Betelgeuse had a brightness of $B\sim2.5$ mag on the night of the Gemini observations (Figure \ref{fig:B-LC}). The star HR 2075 was again used as the PSF standard star. We obtained 6,000 exposures of Betelgeuse, starting at an airmass of 1.15 and ending at an airmass of 1.09. In addition to our standard Fourier analysis of the entire dataset, the 6,000 exposures were split into two independent sets of 3,000 exposures and processed separately to assess the consistency of our results (see Sections \ref{subsec:Diameter} and \ref{subsec:Companion}).

The resulting data products consist of the calculation of statistical magnitude contrast curves, reconstructed images, and astrometric and photometric measurements of any detected companions \citep{howell:2011,horch:2012}, though all the red-channel images were fully saturated because of the brightness of Betelgeuse. Reconstructed images scale the brightest pixel to 1.0 and set the zero level at the mean value of the reconstruction noise, that is, the background reconstructed ``sky" which consists of scatter about a defined n$\sigma$ noise level.
5$\sigma$ is usually the accepted confidence level for a robust companion detection. However, ancillary information can boost the significance of a detection.

\section{Results} \label{sec:results}

In this section we discuss our optical angular diameter measurements of Betelgeuse and the probable direct-imaging detection of its companion.

\subsection{Optical Angular Diameter Measurements of $\alpha$ Ori A}
\label{subsec:Diameter}

The diffraction limit of the Gemini 8-m telescope is 15 mas at 466 nm and 17 mas at 562 nm. As such, Betelgeuse is fully resolved in our speckle images. To measure the effective angular diameter of Betelgeuse in our images, we employed a technique previously used to measure the diameters of Pluto and Charon \citep{Pluto2012PASP..124.1124H}. This technique uses the inflection points (first derivatives) in bisecting cuts through the reconstructed, resolved Betelgeuse images. The absolute value of the derivative along such a cut gives two local maxima, one on each edge of the assumed uniform stellar disk. A measurement of the separation of these two inflection points gives an estimate of the diameter at each wavelength. The assumption of a uniform stellar disk is unlikely to be correct, because there is evidence of non-uniformities on the stellar surface \citep{Kervella2018A&A...609A..67K}, and because limb darkening is expected to be significant at these short wavelengths.  However, the assumption of a uniform disk is appropriate when the signal-to-noise (S/N) and angular resolution are not high enough to effectively map out the surface structure.

\begin{figure*}[h]
\centering
\includegraphics[width=\textwidth]{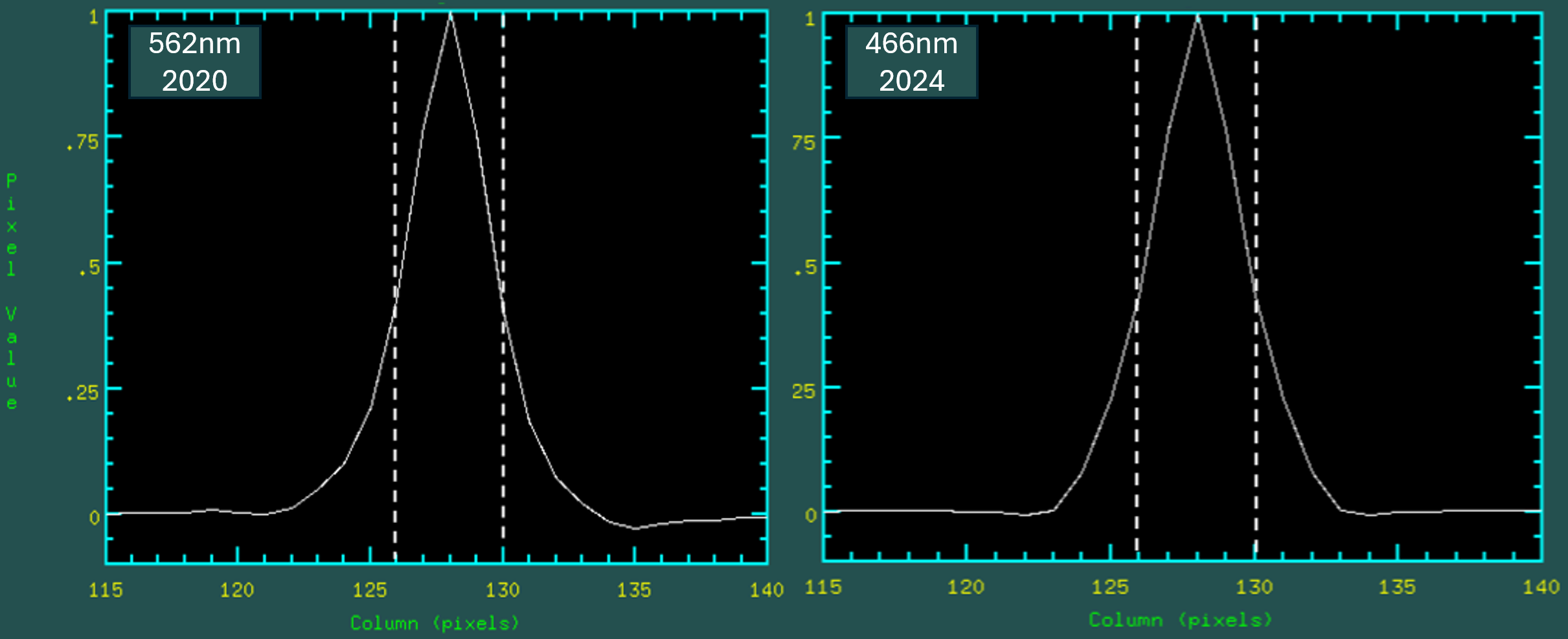}
\caption{Example cuts through the stellar disk 2-D profile of Betelgeuse during the February 2020 and December 2025 observations. The vertical dotted lines show the locations of the first derivative inflection points in each profile (see text). The angular diameters measured in this manner are listed in Table 1.}
\label{fig:diameter}
\centering
\end{figure*}

Figure \ref{fig:diameter} shows the bisecting cuts through the images from February 2020 (at 562 nm) and December 2024 (at 466 nm). The vertical dashed lines show the locations of the inflection points in the first derivative of the stellar profiles. Table \ref{tab:diameter} lists our angular diameter measurements with errors determined by averaging cuts along three orthogonal bisectors. The derived angular diameters for the two observations are in good agreement with each other and with previous angular diameter determinations.

\begin{deluxetable}{ccc}[h]
\tablecaption{Measured Angular Diameters of Betelgeuse\label{tab:diameter}}
\tablehead{
\colhead{UT Date} & \colhead{Wavelength/FWHM} & \colhead{Angular Diameter} \\
\colhead{} & \colhead{(nm)} & \colhead{(mas)}
}
\startdata
2020 February 17 & 562/54 & 40$\pm$1.4 \\
2024 December 09 & 466/44 & 41$\pm$1.2 
\enddata
\end{deluxetable}

We note that our February 2020 observation, and subsequent angular diameter measurement, occurred during the Great Dimming event (Figure \ref{fig:B-LC}), when Betelgeuse was more than a magnitude fainter in the optical than it was during our second observation in December 2024. The nearly equal diameters obtained in February 2020 and in December 2024 suggest that the angular size of the star did not change appreciably during the Great Dimming event. This result provides additional evidence that the Great Dimming event was not the result of a large systemic pulsational event, in which the star significantly changed its radius, but rather the dimming resulted from the dust cloud that formed from ejected photospheric material.

\subsection{The Probable Direct-Imaging Detection of $\alpha$ Ori B}
\label{subsec:Companion}

During the Great Dimming, the companion was eclipsed by Betelgeuse. As such, the companion should not have been visible in our 2020 speckle images, and indeed, they show no sign of a companion.

\begin{figure*}[h]
\centering
\includegraphics[width=0.8\textwidth]{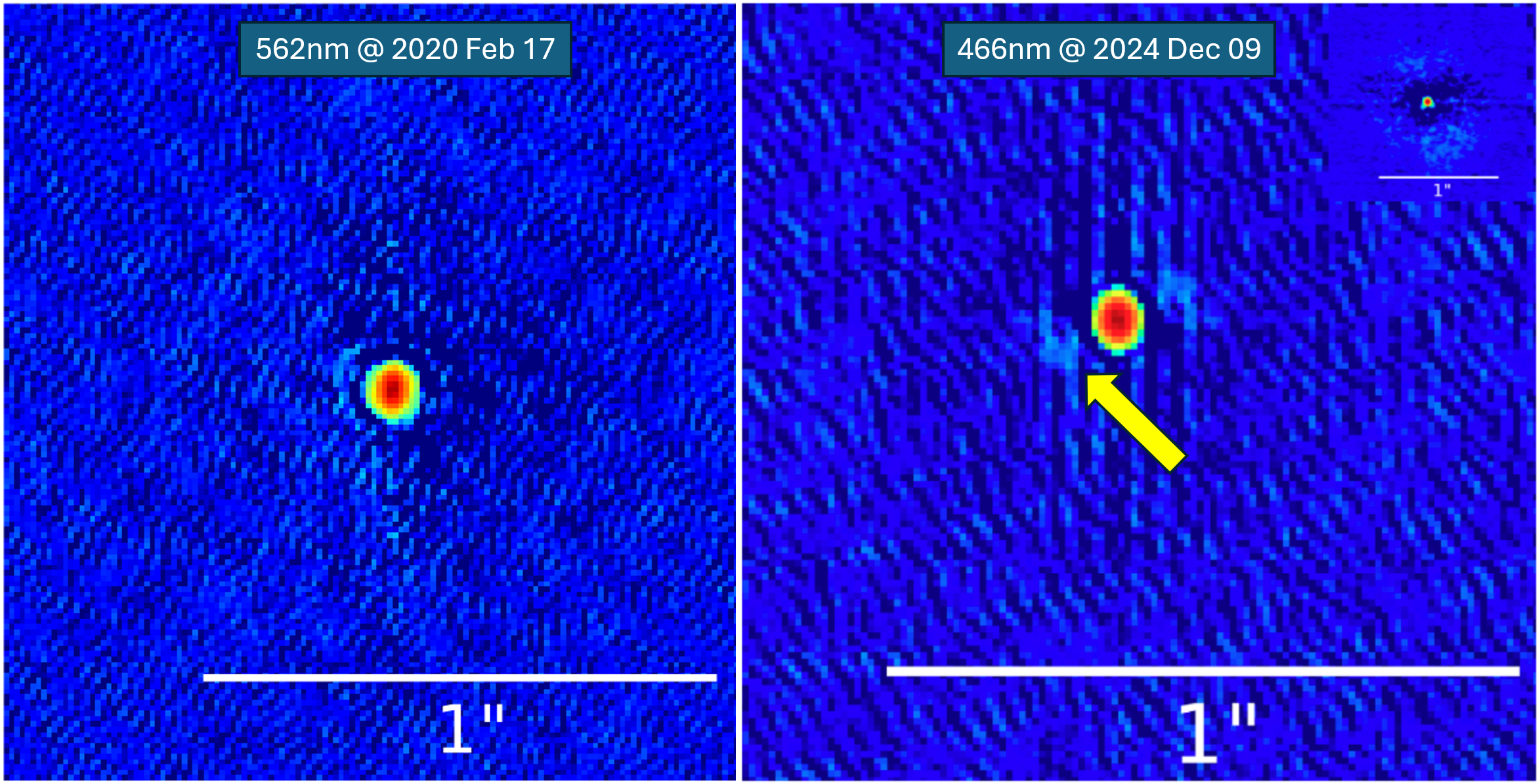}
\includegraphics[width=0.1\textwidth]{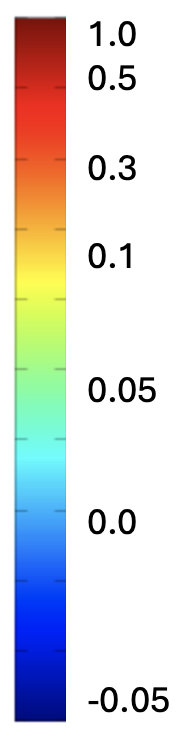}
\caption{The reconstructed images from the 2020 February 17 observations at 562 nm (left) and for the 2024 December 09 observations at 466 nm (right). The images are oriented such that north is up and east is left and the color bar shows the relative scaling applied to both images. The inset image shows a diffraction limited PSF standard star observed near in time to Betelgeuse. Note the probable direct-imaging detection of the companion to Betelgeuse in the 2024 image (arrow) that is absent in the 2020 observations.
\label{fig:Contrast}}
\centering
\end{figure*}

\begin{figure*}[h]
\centering
\includegraphics[width=0.8\textwidth]{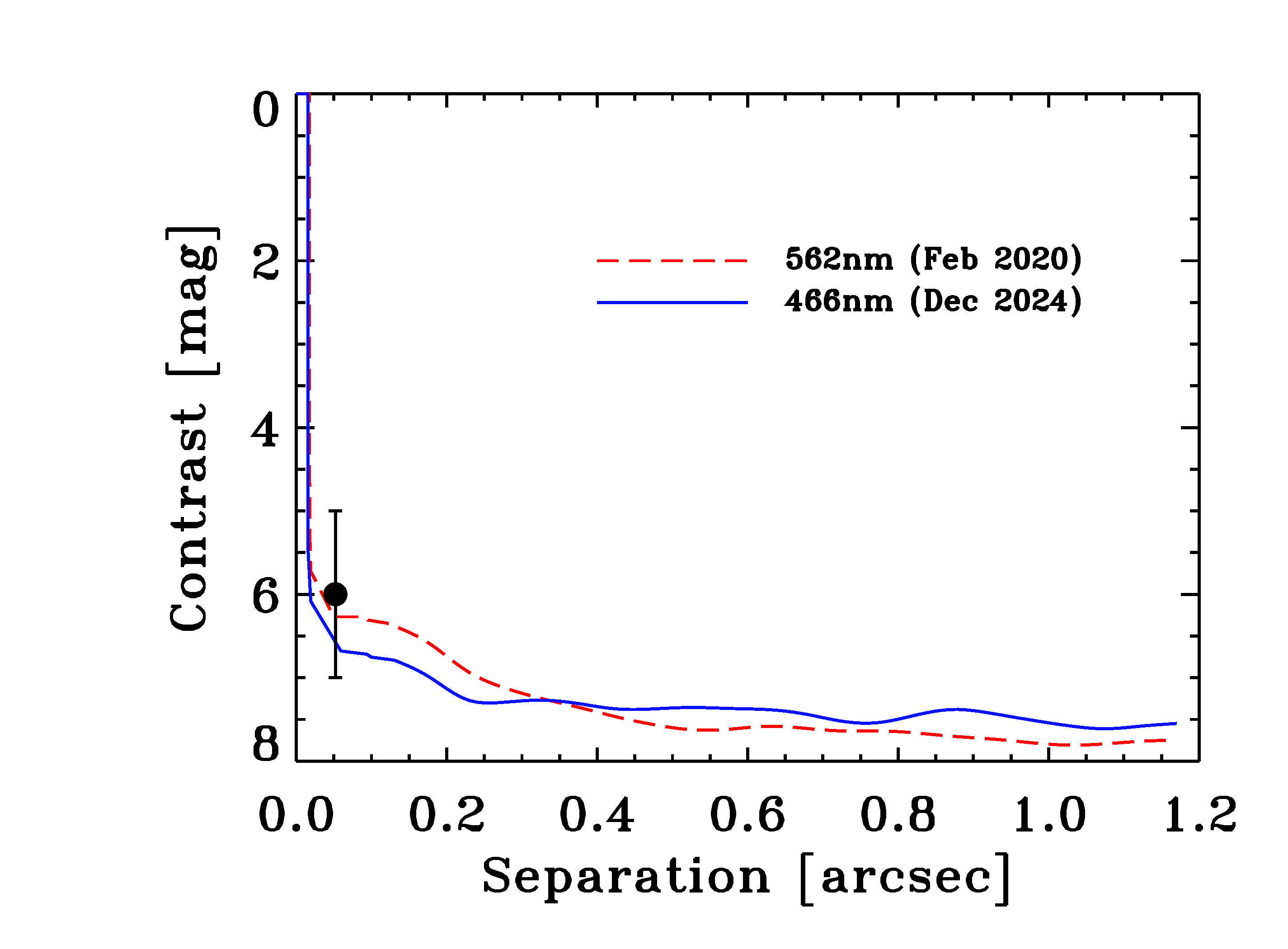}
\caption{The magnitude contrast sensitivity of the February 2020 (dashed red) and December 2024 (solid blue) data plotted as a function of the radial distance from the primary star. The curves represent the sensitivity to a companion at 1$\sigma$. The overplotted point represents the detected companion with 1$\sigma$ uncertainties on its magnitude. The uncertainty on the separation is represented by the size of the plotting symbol. The companion was detected at a significance of $\sim1.5\sigma$.
\label{fig:1sigma}}
\centering
\end{figure*}

However, the proposed stellar companion was predicted to have maximum elongation in its orbit on 2024 December 06, with a predicted angular separation of 52 mas. Our Gemini speckle observations were made on 09 December 2024, only three days after maximum elongation, and indeed, they show a probable direct-imaging detection of a companion to Betelgeuse at the expected location (Figure \ref{fig:Contrast}). The right side of Figure \ref{fig:Contrast} shows a $180^{\circ}$ quadrant ambiguity in the position of the companion, which can occur when taking the autocorrelation of an image. Bispectral analysis of the phase information contained in the images \citep{horch:2011a} can be used to select the correct quadrant; in this case it favors the southeast location.
The differences in the low-level background patterns of the two observations are due to differences in the reconstructed sky in each image, which is unique to every speckle image based on the native seeing, sky brightness, and wavelength of observation. Figure \ref{fig:1sigma} presents our magnitude contrast curves for both the February 2020 and December 2024 observations. These $1\sigma$ sensitivity curves show that the companion was detected at $\sim1.5\sigma$ with a $S/N \sim1.6$. 

To confirm that the detection was not dominated by a hot pixel, cosmic ray, or atmospheric refraction effects, the 2024 data were split into two sets of 3,000 exposures and processed in the same manner as the full set of 6,000 exposures. The full set and each of the two smaller sets show the same hint of a close companion.

\begin{figure*}[h]
\centering
\includegraphics[width=0.6\textwidth]{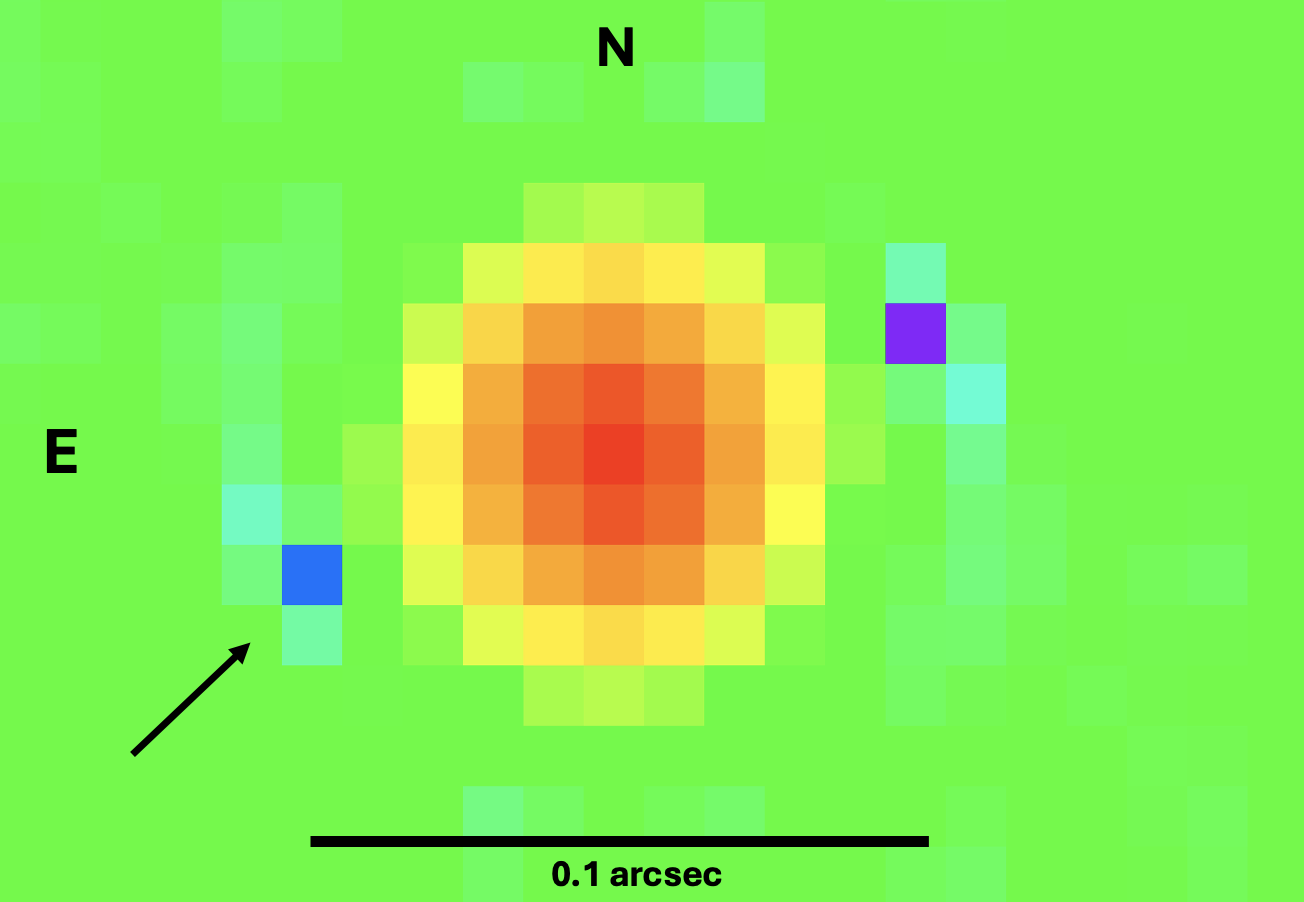}
\includegraphics[width=0.1\textwidth]{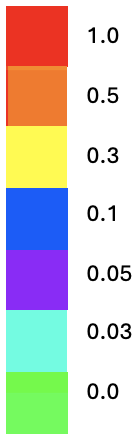}
\caption{A highly stretched version of the 466-nm Betelgeuse image from UT 2024 December 09.
The color scale is arbitrary and the image was used to determine the astrometric position of the companion relative to Betelgeuse. We find a separation of 52 mas and a position angle of 115$^\circ$ degrees. The arrow points to the companion.
\label{fig:reductions}}
\centering
\end{figure*}

To better estimate the astrometric position (separation and position angle) of the detection, we created a severely stretched version of the December 2024 reconstruction image (Figure \ref{fig:reductions}) in order to measure a centroid position of the companion relative to the centroid position of Betelgeuse.
In this image, we have set the zero level at the 1$\sigma$ sensitivity limit shown in Figure \ref{fig:1sigma}. The separation ($52\pm2$ mas) and position angle ($115^\circ\pm5^\circ$) we measure are in excellent agreement with the predictions from \citet{Gold2024ApJ...977...35G}, \citet{MAC2025ApJ...978...50M}, and \citet{Goldberg2025arXiv250518375G}. 

Estimating the brightness of the companion is more difficult. Typically, the magnitude difference of a detected companion is determined in the Fourier plane by measuring the amplitude of the fringes produced by the interference of the primary and companion stars. In this case, the detection is barely above the noise floor, making measurement of the fringe amplitudes more difficult and uncertain. Instead, we estimate the relative flux of the companion from the reconstructed image. We find an estimated peak flux of $\sim0.004\pm0.003$ relative to that of Betelgeuse, which corresponds to a magnitude difference of approximately $\sim6\pm1$ mag.

\section{Discussion}
\label{sec:discussion}

In this section we discuss the likelihood that our detection is real and bound to Betelgeuse, as well as the nature of the companion.

\subsection{Likelihood That the Companion Is Real and Bound}

Several recent studies have concluded that a stellar companion is the most likely cause of the observed LSP photometric variations of Betelgeuse, and have investigated its nature. \citet{Gold2024ApJ...977...35G} proposed a companion with a semi-major axis of 1850 R$_\odot$ (2.43 $R_{Betelgeuse}$) and $M_\star\sin i = 1.17 M_\odot$. With a distance of 168 pc \citep{Joyce2020ApJ...902...63J}, the maximum separation between Betelgeuse and its companion is $\sim0.06\arcsec$. The predicted maximum angular separations at the radial velocity quadrature points on 2024 December 06, 2027 November 26, 2030 November 15, and 2033 November 04. \citet{MAC2025ApJ...978...50M}, independently, also presented evidence for a companion to Betelgeuse. They find a semi-major axis of 1818 R$_\odot$ and, adopting a mass for Betelgeuse of $\sim 17.5 M_\odot$, a mass of $\le 1.25 M_\odot$, with an RV-fit preferred value of $0.6M_\odot$. Follow-up papers used the Hubble Space Telescope and the Chandra X-ray Observatory \citep{Goldberg2025arXiv250518375G, OGrady2025arXiv250518376O} to further refine the nature and mass of the companion star. In our speckle images of Betelgeuse, we find a probably direct-imaging detection of this same companion star.

In a blind search for close stellar companions to bright stars, the results presented here would be marginal, but the correspondence between our detection and the predicted location of the companion adds credence to the result. The companion is not detected in 2020 -- as expected -- and is detected in 2024 at the separation and position angle predicted by \citet{Gold2024ApJ...977...35G}, \citet{MAC2025ApJ...978...50M}, and \citet{Goldberg2025arXiv250518375G}. In the context of these other works, we regard the results presented here as more robust than if they were presented on their own. Nonetheless, there are several other explanations for this detection that should be addressed.

\begin{figure*}[h]
\centering
\includegraphics[width=\textwidth]{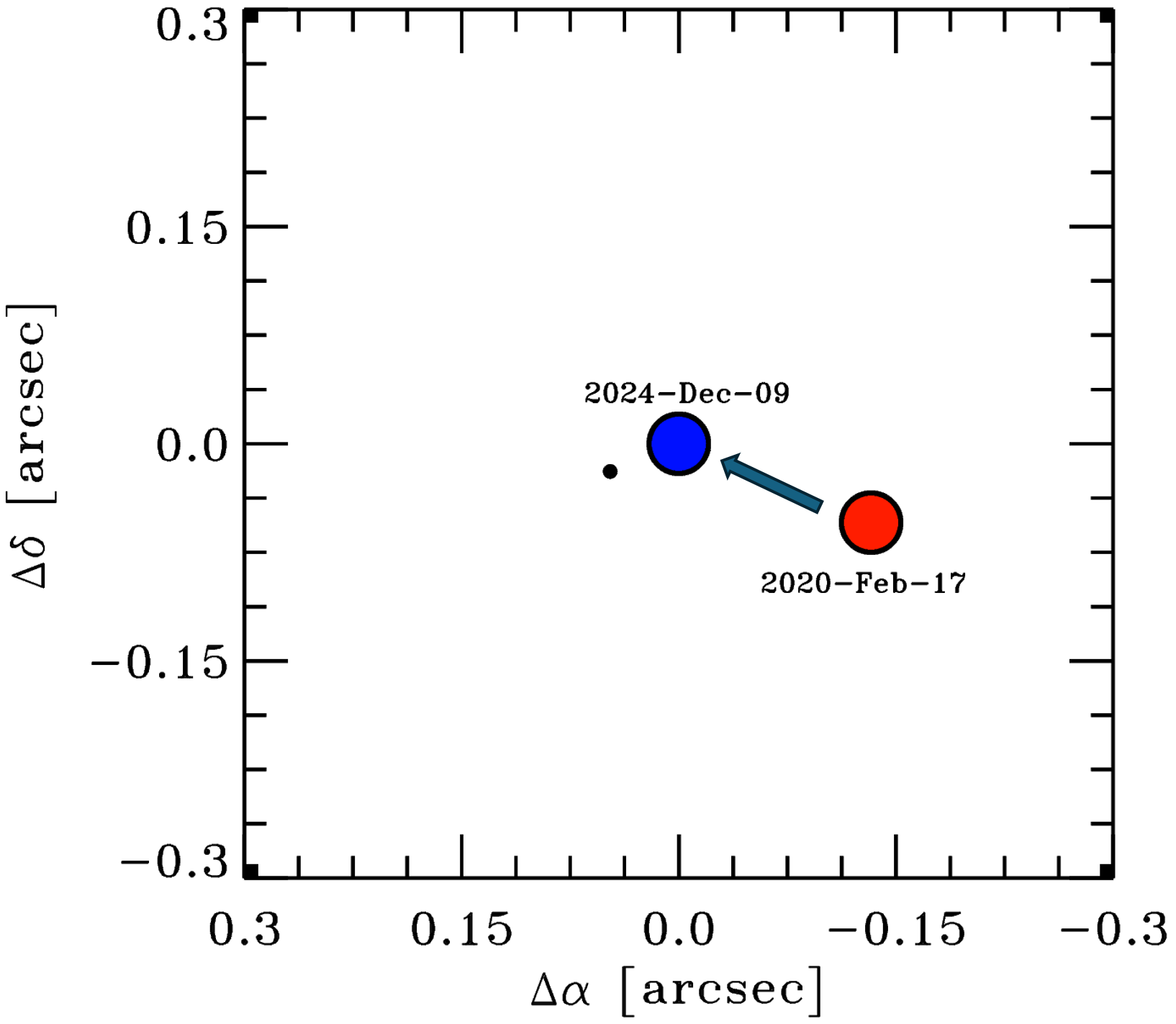}
\caption{Schematic diagram of the on-sky proper motion of Betelgeuse from 2020 (red) to 2024 (blue). [0,0] is centered on the 2024 position of Betelgeuse, and the size of the circles represent the 40-mas angular diameter of Betelgeuse. The black dot shows the relative position of the detection to the 2024 position of Betelgeuse. This diagram shows that if the detection were a background star, it would have been detectable in our 2020 speckle observations.
\label{fig:pm}}
\centering
\end{figure*}

The first is that the detection is a background star unrelated to Betelgeuse. Betelgeuse has a proper motion of $\mu_\alpha = 27.54\pm1.03$ mas/yr and $\mu_\delta = 11.30\pm0.51$ mas/yr \citep{vanLeeuwen2007}, so over four years, Betelgeuse moved $\sim 0.12\arcsec$ to the northeast (Figure \ref{fig:pm}). If the companion were a background star, it would have been separated from Betelgeuse by $\sim0.2\arcsec$ in 2020 and easily detectable in our speckle images, but we did not detect a companion. As such, the companion is most likely not a star lurking in the background.

Second, it is possible that the detection is a foreground star with a very high proper motion. While the formal field-of-view of the speckle cameras is approximately 6\arcsec, the coherent field-of view-that is useful for speckle interferometry is approximately 1-1.2\arcsec\ in radius from the primary star.  Therefore, a foreground star would have to move $\gtrsim 1\arcsec$, from outside the field-of-view to nearly the center of the array in four years. This would correspond to a proper motion of $\mu \gtrsim 250$ mas/yr. Stars with such large proper motions are typically within $\sim 30$ pc of the Sun \citep{hipparcos1997}. With an apparent magnitude of $m_B \approx 8$mag, the absolute $B$ magnitude of the detection would be $M_B \sim 10-11$ mag, corresponding to a late-K or early-M dwarf \citep{pm2013}. While not impossible, it is unlikely that such a bright, nearby, fast-moving star would have remained hidden in the glare of Betelgeuse for centuries, and that the proper motion vector of the star placed it at the exact location of the predicted companion to Betelgeuse. As such, we consider this scenario rather contrived, and instead, that our detection is the predicted companion to Betelgeuse.

\subsection{Nature of the Companion}

Given the properties of our detection, what could it be? The 466-nm narrowband filter used to observe the companion has a central wavelength similar to that of the standard $B$-filter (445 nm). At the time of the December 2024 observations, the $B$-magnitude of Betelgeuse was approximately $B\approx2.5\pm0.1$ mag (AAVSO and Figure \ref{fig:B-LC}). The apparent magnitude of the companion was therefore $B\approx8.5\pm1$ mag.  At a distance of 168 pc, Betelgeuse has an absolute magnitude of $M_B \approx -3.6$ mag, so the companion would have an absolute magnitude of $M_B \approx 2.4\pm1$ mag.

\begin{figure*}[h]
\centering
\includegraphics[width=0.9\textwidth]{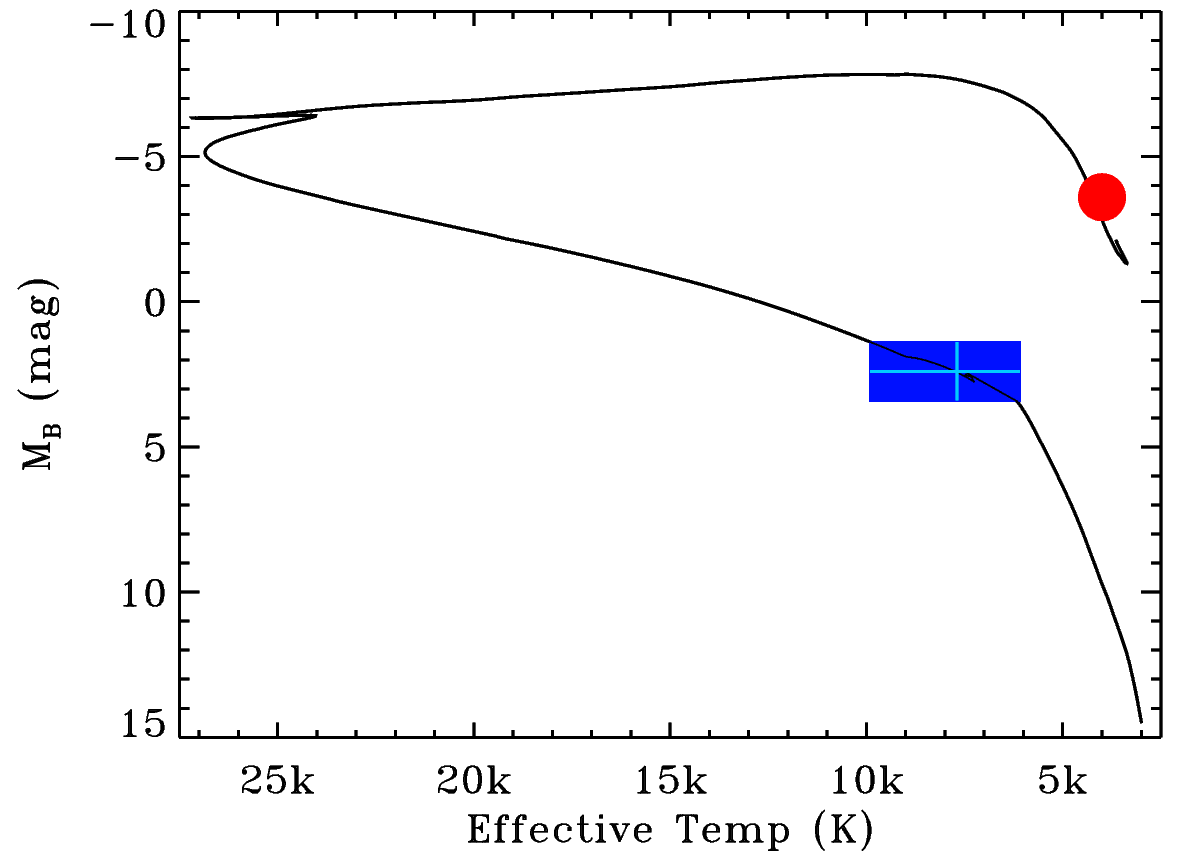}
\caption{Absolute $B$ magnitude versus effective temperature for a solar-metallicity, 10-Myr MIST isochrone. The red dot marks the approximate location of Betelgeuse, and the blue box represents the possible boundaries for the stellar companion based on the estimated absolute $B$ magnitude, the estimated uncertainty on that $B$ magnitude, and where the absolute $B$ magnitude boundaries intersect the isochrone. These values are in agreement with other works that predicted the presence of a companion \citep{Gold2024ApJ...977...35G, MAC2025ApJ...978...50M}.
\label{fig:isochrone}}
\centering
\end{figure*}

Assuming the two stars are coeval and have an age of 10 Myr, they can be placed on a single-age isochrone to estimate the possible mass range of the stellar companion from its absolute $B$ magnitude. Using the MIST stellar evolution models \citep{mist0,mist1,mist2, mist3, mist4, mist5}, we generate a 10-Myr solar metallicity stellar evolution isochrone for the system (Figure \ref{fig:isochrone}). On the isochrone, Betelgeuse has a mass of $\sim 20 M_\odot$, which is in agreement with the other works. At an absolute magnitude of $M_B\sim2.4$ mag, the companion has an effective temperature and mass of $T_{eff} \sim 7400$ K and $M \sim 1.6 M_\odot$, respectively. The absolute $B$ magnitude of the companion is uncertain by at least one magnitude, so the companion could be as massive as $\sim 2 M_\odot\ (T_{eff} \sim 10,000$ K) or as small as $\sim 1.4 M_\odot\ (T_{eff} \sim 6000$ K).

Given the low S/N detection of the companion and the relatively poor constraints on its brightness, it should be noted that the physical constraints on the properties of the companion are extremely loose. With a detection in only one filter, there is no information on the color-temperature of the star. Additionally, there has been no accounting for the fact that the companion may be affected by the evolution of Betelgeuse itself. It may orbit within the envelope and ejecta of Betelgeuse, or its formation and evolution may be stunted by the presence of the supergiant. However, the results are in general agreement with the mass estimates that were based on dynamical considerations \citep{Gold2024ApJ...977...35G, MAC2025ApJ...978...50M}.

Other recent works suggest that the companion may be lower-mass ($\sim 0.6 M_\odot$) based on the limits of FUV and X-ray emission expected from 10-Myr, Solar-like stars \citep{Goldberg2025arXiv250518375G, OGrady2025arXiv250518376O}, which is in disagreement with our analysis. But a 10-Myr, $0.4-0.6 M_\odot$ star would be $M_B\sim9$ - $M_B\gtrsim13$ magnitudes fainter than Betelgeuse, which is beyond the reach of speckle imaging on an 8.1-meter telescope. 
The fact that we detected a companion, and the observed magnitude difference between Betelgeuse and this companion, suggests that it is approximately solar-mass in size and evolving toward the zero-age main sequence. However, it will likely never arrive at that stage, as Betelgeuse is predicted to produce a much-anticipated supernova in the coming millennia.

\section{Summary and Future Work}
\label{sec:summary}

Recent work has proposed that the long-period photometric variation and the similarly phased astrometric and radial velocity variations observed in Betelgeuse are caused by a $\sim 1-2 M_\odot$ stellar companion. The stellar companion was predicted to be at greatest elongation from Betelgeuse in December 2024, with a separation of tens of mas.

We observed Betelgeuse during the Great Dimming event in 2020 and three days after the predicted date of greatest elongation with the optical speckle imaging camera `Alopeke on the 8.1-meter Gemini North telescope.

In both observations, the stellar disk of Betelgeuse is spatially resolved. The angular diameter, and thus the physical stellar radius, is consistent with previous measurements, and does not change significantly between 2020 and 2024. These observations provide additional evidence that the Great Dimming event was not caused by a rapid and significant change in the physical radius of the star, but rather a large ejected dust cloud.

The 2024 observations reveal the probable direct-imaging detection of the companion to Betelgeuse. This is in agreement with the expectation that in 2020 the companion would be ``behind'' Betelgeuse and unobservable, and that in 2024, the companion would be at its greatest elongation. Our detection is $\sim$6 magnitudes fainter than Betelgeuse in the $B$-band with an angular separation of 52 mas from Betelgeuse, in excellent agreement with recent works. Assuming the companion is coeval with Betelgeuse with an age of 10 Myr, the companion has a mass of $\sim 1.6 M_{\odot}$ and is likely a young, pre-main-sequence F-dwarf.

The results presented here are not definitive, as the detection is at the limit of the instrument capabilities. However, the results do present the most direct and substantive evidence for the existence of a stellar companion to Betelgeuse, as well as the properties of that companion. The next predicted greatest elongation for the stellar companion is UT 2027 November 26 \citep{Gold2024ApJ...977...35G}; we recommend that the community observe Betelgeuse prior to and during that event to better constrain the nature of the companion.

The name Betelgeuse means ``Hand of the Giant,'' with ``Elgeuse'' being a historical Arabic name of the Orion constellation and a feminine name in old Arabian legend. Given that $\alpha$ Ori B orbits the hand of the giant, we suggest that the companion star be named Siwarha, or ``Her Bracelet.''

\vspace{0.25in}
We would like to thank the referee for a timely and productive review. The authors thank Salma Bejaoui for her Arabic translation effort and providing the suggested name for its companion.  The observations in this paper -Gemini proposal GN-2024B-FT-111 - made use of the high-resolution imaging instrument `Alopeke. `Alopeke was funded by the NASA Exoplanet Exploration Program and built at the NASA Ames Research Center by Steve B. Howell, Nic Scott, Elliott P.~Horch, and Emmett Quigley. `Alopeke was mounted on the Gemini North 8-m telescope of the international Gemini Observatory, a program of NSF NOIRLab, which is managed by the Association of Universities for Research in Astronomy (AURA) under a cooperative agreement with the U.S. National Science Foundation, on behalf of the Gemini partnership: the National Science Foundation (United States), National Research Council (Canada), Agencia Nacional de Investigación y Desarrollo (Chile), Ministerio de Ciencia, Tecnología e Innovación (Argentina), Ministério da Ciência, Tecnologia, Inovações e Comunicações (Brazil), and Korea Astronomy and Space Science Institute (Republic of Korea).  This work was enabled by observations made from the Gemini North telescope, located within the Maunakea Science Reserve and adjacent to the summit of Maunakea. We are grateful for the privilege of observing the Universe from a place that is unique in both its astronomical quality and its cultural significance.  This research has made use of the Exoplanet Follow-up Observation Program (ExoFOP; DOI: 10.26134/ExoFOP5) website, which is operated by the California Institute of Technology, under contract with the National Aeronautics and Space Administration under the Exoplanet Exploration Program.  D.R.C. acknowledges support from NASA through the XRP grant No.~18-2XRP18\_2-0007.

\noindent {\it Facilities:} Gemini -  `Alopeke

\bibliographystyle{aasjournal.bst}
\bibliography{betelgeuse}

\end{document}